\documentclass[11pt]{article}
\usepackage{amsmath}
\usepackage{jheppub}
\usepackage{amssymb}
\usepackage{graphicx}
\usepackage[T1]{fontenc}
\usepackage{slashed}
\usepackage{hyperref}
\usepackage[utf8]{inputenc}
\usepackage{xspace}
\usepackage{color}
\usepackage{ulem}
\usepackage{array}
\usepackage{ulem,fancyvrb}
\usepackage{xcolor}
\usepackage{mathtools}

\allowdisplaybreaks 

\date{\today}

\date{\today}
\begin{document}

\title{Exploring the Impact of Dissipation Coefficient in Warm Higgs Inflation}

\author[a,1]{Wei Cheng,}
\affiliation[a]{School of Science, Chongqing University of Posts and Telecommunications, Chongqing 400065, China}
\emailAdd{chengwei@cqupt.edu.cn}
\author[b,1]{Xue-Wen Chen,}
\emailAdd{chenxuewen@cqust.edu.cn}
\author[a,1]{Ruiyu Zhou,}
\emailAdd{zhoury@cqupt.edu.cn}
\author[a]{Jiu-Jiang Jiang,}
\author[a]{Xin-Rui Dai,}
\author[a]{Zi-Han Zhang,}
\author[a]{and Tong Qin}

\affiliation[b]{Chongqing University of Science and Technology, Chongqing 401331, China}

\footnotetext[1]{Corresponding authors.}

\abstract{
In this study, we conducted a detailed analysis of the core parameter of Warm Higgs Inflation (WHI) —the dissipation coefficient ($Q$). As a crucial parameter in the warm inflation process, $Q$ exerts profound influences on the entire evolutionary process. By meticulously deriving the relationships between various quantities and $Q$, we successfully circumvented the common preconceptions regarding strong and weak dissipation, laying the foundation for a more accurate exploration of their interconnections. Taking into account the constraints imposed by Cosmic Microwave Background, we observed that the dissipation coefficient $Q$ remains at extremely low levels throughout the entire warm inflation process, i.e., $Q \ll 1$. This observation indicates that WHI falls under the category of weakly dissipative warm inflation. Despite being weakly dissipative, $Q$ still plays a crucial role in the evolution of temperature, energy, and other quantities, highlighting its significance and non-negligibility. We delved deeper into the impact of the primordial power spectrum on the dissipation coefficient $Q$ during the warm inflation process, discovering that the dependency is not significant. Consequently, this naturally leads to the unobtrusive dependence of the gravitational wave power spectrum on $Q$. Finally, we found that gravitational waves generated by WHI hold the potential for verification in future observational experiments, especially through the SKA100 experiment. These findings provide a theoretical support for a more profound understanding of the early evolution of the universe.
}

\maketitle
\preprint{}

\section{Introduction}
Inflationary cosmology has emerged as the pivotal framework for understanding the dynamics of the early universe, addressing fundamental issues such as the flatness, and the horizon problems~\cite{Guth:1980zm,Linde:1981mu,Linde:1983gd} while providing a mechanism to explain the inhomogeneities observed in the Cosmic Microwave Background Radiation (CMB)~\cite{AlHallak:2022haa}. Within this cosmological paradigm, two principal models of inflation have been distinguished: cold inflation~\cite{Cheng:2018ajh,Cheng:2018axr,Cheng:2022hcm,Cheng:2022pzs,Khan:2023uii,Rodrigues:2023xqu,Zhou:2022ovp,Oikonomou:2022pdf,Kannike:2018zwn,Ema:2017rqn,Lebedev:2011aq,Calmet:2016fsr,Bezrukov:2007ep} and warm inflation~\cite{Li:2018wno,Li:2019zbk,Taylor:2000ze,Berera:1996fm,Berera:1999ws,Bastero-Gil:2018uep,Bastero-Gil:2009sdq,Levy:2020zfo,Benetti:2016jhf}.

Although the cold inflation model has achieved significant success in depicting the exponential expansion of the universe and the generation of density perturbations due to quantum fluctuations of the inflation field, it falls short in explaining the transition from inflationary to radiation dominated era. In contrast, the warm inflation model introduces a novel mechanism that maintains thermal equilibrium and continuous production of relativistic particles, offering a more seamless transition from the inflationary phase to a radiation-dominated stage~\cite{Berera:1995ie,Berera:1995wh,Cheng:2021qmc,Cheng:2021nyo}. A recent review of warm inflation has been discussed in Refs.~\cite{Kamali:2023lzq,Gron:2016nxb}.

Among the plethora of models for warm inflation, the Warm Higgs Inflation (WHI) has garnered extensive attention in the scientific community~\cite{Kamali:2018ylz,Samart:2021eph,Eadkhong:2023ozb}. This model not only incorporates the unique properties of the Higgs field but also explores its complex interactions with the inflationary process and dissipative phenomena. Given the pivotal role of the Higgs field in the Standard Model, especially in imparting mass to fundamental particles, its involvement in the warm inflation framework adds complexity and depth to our understanding of the early universe's dynamics.

This study focuses on a detailed analysis of a critical parameter–dissipation coefficient denoted as $Q$ – within the framework of the WHI. We aim to transcend the conventional dichotomy of strong ($Q \gg 1$) and weak ($Q \ll 1$) dissipation~\cite{Kamali:2018ylz,Samart:2021eph,Eadkhong:2023ozb}, adopting a more nuanced approach to unveil the subtle effects of $Q$ within the context of warm inflation. Specifically, this study will explore the dynamical characteristics of $Q$ and its impact on the early universe's temperature, energy, and other key physical quantities, as well as its intricate relationship with the primordial power spectrum.

Moreover, this research will investigate the dependency of the gravitational wave power spectrum on $Q$ during WHI, particularly considering the verification prospects in future observational experiments, such as the SKA100 project~\cite{Carilli:2004nx}. The goal of this study is to provide a comprehensive theoretical foundation for understanding the physical processes of the early universe's evolution under the WHI framework and to offer theoretical guidance for detecting gravitational waves and other cosmological phenomena in the early universe.

The rest parts are arranged as follows
In Sec.~\ref{sec:WSFI}, we will commence by providing a concise overview of the background evolution of WHI. Subsequently, we will delve into a comprehensive exposition of the $Q$-dependent WHI and explore the $Q$-dependent relic gravitational waves.
Sec.~\ref{sec:Num} will conduct a numerical discussion, encompassing the dynamic evolution of various quantities throughout WHI process. We will also investigate the relationship between the spectral index and tensor-to-scalar ratio, along with the $Q$-dependent relic gravitational waves.
Finally, we will briefly summarize in Sec.~\ref{sec:Sum}.

\section{Dissipation coefficient $Q$}\label{sec:WSFI}

\subsection{Background evolution of warm Higgs inflation}
This section provides a detailed exposition of the background evolution of WHI, aiming to comprehensively understand the dynamics of the universe during this crucial period and laying the foundation for the subsequent examination of the dissipation coefficient's influence. The evolution of various energy components in the universe is governed by the Friedmann equation. During the WHI phase, the cosmic energy predominantly consists of the energy density of the Higgs inflaton field, $\rho_{h}(t)$, and the radiation energy density, $\rho_{r}(t)$. Considering the Einstein frame action with the flat FLRW line element, the Friedmann equation for WHI can be succinctly formulated as:
\begin{eqnarray}
H^2 =\frac{1}{3\,M_p^2}\left( \rho_{h} + \rho_r\right),\label{E1}
\end{eqnarray}
where the evolution equations for the homogeneous Higgs inflaton field energy ($\rho_h$) and the radiation energy ($\rho_r$) can be derived by separately analyzing the corresponding field motion equations. Specifically, in the context of WHI, their explicit forms are given by:
\begin{eqnarray}
\dot \rho_h + 3 H ( \rho_h + p_h) &=& - \Gamma ( \rho_h +p_h) \,,\label{rhoh} 
\\
\dot \rho_r + 4 H \rho_r  &=& \Gamma \dot{h}^2\,. \label{rhor}
\end{eqnarray}
where the pressure $p_h$ is defined as $p_h=\dot h^2/2 - U(h)$, and $\rho_h + p_h= \dot h^2$. The general form of the $\Gamma$ is given by~\cite{Berera:1998gx,Zhang:2009ge,Bastero-Gil:2011rva,Berera:2001gs}
\begin{eqnarray}
\Gamma  = C_{m}\frac{T^{m}}{h^{m-1}}\,, \label{Q1}
\end{eqnarray}
where $m$ is an integer and $C_{m}$ is associated to the dissipative microscopic dynamics which is a measure of inflaton dissipation into radiation. Different choices of $m$ yield different physical descriptions, e.g., Refs.\cite{Zhang:2009ge,Berera:2001gs,Bastero-Gil:2011rva}. In this study, we will set $m=1$ for our calculations in consideration of the elevated temperature, as discussed in~\cite{Berera:2008ar}.


In slow-roll scenario, Eqs.~(\ref{rhoh}) and (\ref{rhor}) can be further reduced to:
\begin{eqnarray}
3 H ( 1 + Q ) \dot h &\simeq&  - U_h    \,,\label{eominfh} \\
4 \rho_r  &\simeq& 3 Q\dot h^2\,. \label{eomradsl}
\end{eqnarray}
where $Q$ is the dissipative coefficient with $Q=\Gamma/(3 H)$.

\subsection{$Q$-dependent warm Higgs inflation}

In this subsection, we meticulously analysis the impact of the dissipation coefficient on the WHI.
We start with the action of the Standard Model Higgs doublet $H$ with a non-minimal coupling to gravity in the Jordan  frame:
\begin{eqnarray}\label{starting action}
    S_{J}=\int d^{4}x\sqrt{-g_J}\left[\frac{M_{p}^2}{2}\left(1+2\xi\frac{H^{\dagger} H}{M_{p}^2}\right)R_J+g_J^{\mu\nu}(D_\mu H)^\dag (D_\nu H)-V(H)\right],
\end{eqnarray}
where $M_p$ is the Planck mass, $\xi$ is a coupling constant, $R_J$ is the Ricci scalar, and $V(H)$ is the potential of the Higgs field, neglecting the mass term. This choice is motivated by the understanding that the impact of the quadratic term relative to the quartic term on inflation is negligible during the inflationary process.

In the unitary gauge, employing the conformal transformation facilitates the conversion of the potential $V(H)$ to the Einstein frame, as below:
\begin{eqnarray}\label{potential}
\quad U&\simeq& \frac{\lambda M_p^4}{4\xi^2}\left(1+\mathrm{exp}\left(-\sqrt{\frac{2}{3}}\frac{h}{M_p}\right)\right)^{-2}.
\end{eqnarray}
where $\lambda$ is the self-coupling of the Higgs doublet.

In a thorough investigation of the slow-roll inflation process, through a comprehensive analysis of the inflation potential, we have obtained the slow-roll parameters ($\varepsilon$ and $\eta$) and ($\varepsilon_{H}$ and $\eta_{H}$), respectively corresponding to cold inflation and warm inflation:
\begin{eqnarray}
\varepsilon &=& \frac{M_p^2}{2}\left( \frac{U'}{U}\right)^2\,,\quad \eta = M_p^2\,\frac{U''}{U}\,;
\label{SR-parameters}\\
\varepsilon_{H} &=&\frac{\varepsilon}{1+Q}\,,~~~~~\quad \eta_{H} = \frac{\eta}{1+Q}\,.
\label{slowroll}
\end{eqnarray}

Furthermore, we can delineate the characteristics of the universe during inflation, i.e., ($\varepsilon \ll 1 $ and  $\eta \ll 1 $) as well as ($\varepsilon_H \ll 1$ and  $\eta_H \ll 1$), which 
also ensures the flatness of the inflation potential. Once these conditions are violated, i.e.,($\varepsilon \approx 1 $ or  $\eta \approx 1 $) as well as ($\varepsilon_H \approx 1$ or  $\eta_H \approx 1$), the slow-roll process comes to an end, providing a framework for determining the field values at the conclusion of inflation.

In contrast to the cold scenario, the dissipative effects resulting from warm expansion will impact the scalar power spectrum. Detailed derivations can be found in the Refs.\cite{Graham:2009bf,Ramos:2013nsa,Visinelli:2016rhn,Moss:2007cv,Hall:2003zp}, and the final expression is as follows: 
\begin{eqnarray}
P_{\cal R}(k) = \left( \frac{H_k^2}{2\pi\dot h_k}\right)^2\cal{F}\,,
\label{spectrum}
\end{eqnarray}
where ${\cal{F}}$ is the modified factor, and its specific form is given by:
\begin{eqnarray}
{\cal{F}} = \left( 1 + 2n_k +\left(\frac{T_k}{H_k}\right)\frac{2\sqrt{3}\,\pi\,Q_k}{\sqrt{3+4\pi\,Q_k}}\right)G(Q_k)\,,
\label{F}
\end{eqnarray}
where the subscript "$k$" denotes the time when the cosmological perturbation mode with wave number "$k$" exits the horizon during inflation. For the convenience of numerical computations in the next section, we establish a connection between each component in the scalar power spectrum and the Higgs inflaton field $h$, as detailed in Appendix~\ref{App}.

This dissipative effect has negligible influence on the tensor power spectrum, which originates from the primordial tensor fluctuations of the metric. Its form aligns with the cold inflation model presented in Refs.\cite{Kinney:2003xf,Dodelson:2003ft} as
\begin{eqnarray}
P_{T}(k) = \frac{16}{\pi}\Big(\frac{H_{k}}{M_{p}}\Big)^{2}
\,. \label{PT}
\end{eqnarray}

Given the significant impact of dissipative effects on the scalar power spectrum during WHI, with negligible effects on the tensor power spectrum, it can be inferred that in this context, the tensor-to-scalar ratio ($r$) obtained through the ratio of tensor power spectrum to scalar power spectrum, and the spectral index ($n_s$) derived from the scalar power spectrum, will be modified. Specifically, these corrections can be summarized as follows~\cite{AlHallak:2022haa}:
\begin{eqnarray}
r &=& \frac{P_{T}(k)}{P_{\cal R}(k)}\Bigg|_{k=k_{p}}=\frac{16\epsilon}{(1+Q)^2}\mathcal{F}^{-1}\,. \label{r}\\
n_{s}-1&=&\frac{d \ln P_{\cal R}(k)}{d\ln k}\Bigg|_{k=k_{p}}=\frac{1}{H P_{\cal R}}\frac{d P_{\cal R}}{dh}\dot{h}\,,\label{ns1}
\end{eqnarray}
where $k_{p}$ corresponds to the pivot scale. 

\subsection{$Q$-dependent of relic gravitational waves}\label{sec:RGW}

This section explores the influence of the dissipation coefficient on the relic gravitational waves (GWs).
In the scenario of WHI, the expression for the energy spectrum of induced GWs during the radiation-dominated era can be formulated through the utilization of the curvature perturbation power spectrum $P_\mathcal{R}(k)$ (Eq.~(\ref{spectrum})) dependent on $Q$ as follows \cite{Kohri:2018awv,Peng:2021zon}:
  \begin{align}\label{IGW}
		\Omega_{\rm{GW}}&(\eta,k) = \frac{1}{6} \int^\infty_0 dv \int^{|1+v|}_{|1-v|}du \left( \frac{4v^2-(1+v^2-u^2)^2}{4uv}\right)^2I^2(u,v)P_\mathcal{R}(ku)P_\mathcal{R}(kv),\nonumber\\
\end{align}
where  
\begin{align}I^2(u,v)=&\frac{1}{2} \left( \frac{3(u^2+v^2-3)}{4u^3v^3}\right)^2 \left\{\left[-4uv+(u^2+v^2-3) \ln\left| \frac{3-(u+v)^2}{3-(u-v)^2}\right| \right]^2 \right. \nonumber\\
&\left.+ \pi^2(u^2+v^2-3)^2\Theta(v+u-\sqrt{3})\right\}.
\end{align}

Taking the thermal history of the Universe into consideration, one can get the GWs spectrum at present~\cite{Kohri:2018awv,Correa:2023whf},
\begin{equation}
	\Omega_{\rm GW,0}(k)=0.39\times \Omega_{\gamma,0} \left(\frac{g_{\star,0}}{g_{\star,\rm eq}}\right)^{1/3}  \Omega_{\rm GW}(\eta_{\rm eq},k),
\end{equation}
where $\Omega_{\gamma,0}$ is the density parameter of radiation today, $g_{\star,0}=3.36$ and $g_{\star,\rm eq}=3.91$ are the effective numbers of relativistic degrees of freedom at the present time and at the time $\eta_{\rm eq}$ of the radiation-matter equality, respectively.

In order to facilitate a comparative analysis with gravitational waves experiments, we transform the variation of gravitational waves with respect to wave number $k$ into a corresponding variation with respect to frequency $f$. The relationship between $k$ and $f$ is delineated as follows~\cite{Correa:2023whf}:
\begin{eqnarray}
f=\frac{k}{2\pi}=1.5\times10^{-15}\bigg(\frac{k}{1~\rm{Mpc}^{-1}}\bigg)~\rm{Hz}.
\end{eqnarray}

\section{Numerical Discussion}\label{sec:Num}

In this section, we begin by constraining the model parameters of WHI using CMB, and then further investigate physical quantities such as $r$, $n_s$ and $T$, along with their dependence on $Q$.

\subsection{Spectral index and tensor-to-scalar ratio}

\begin{figure}[!thb]
  \centering
  \includegraphics[width=0.55\textwidth]{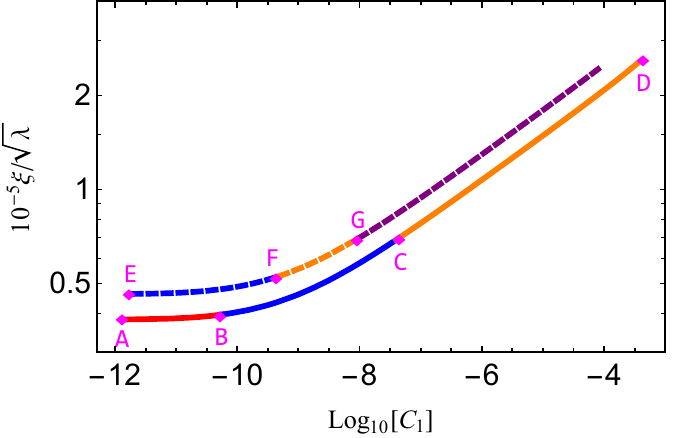}
  \caption{The evolution of $\xi/\sqrt{\lambda}$ with respect to $C_1$ within the framework of the WHI model. The dashed line corresponds to $N_e=60$, while the solid line corresponds to $N_e=50$.}\label{x-C1}
\end{figure}

Fig.~\ref{x-C1} illustrates the viable parameter space of the WHI model under consideration of the amplitude of scalar fluctuations $\Delta_{\mathcal{R}}^2=2.2\times10^{-9}$ constraints~\cite{Planck:2015fie}. The dashed line corresponds to $N_e=60$, whereas the solid line corresponds to $N_e=50$. To enrich the foundation of our forthcoming discussions, for precise values of $C_1$, $\xi/\sqrt{\lambda}$, and $Q$ from Point A to Point G, kindly refer to Tab.~\ref{Points}. Notably, the value of $Q$ corresponds to the end of WHI. In the ensuing discourse, we will scrutinize predictions associated with each of these points individually.

\begin{table}[tb]
\caption{The $(10^{10}C_1, 10^{-4}\xi/\sqrt{\lambda})$ value of the critical points (A, B, C, D, E, F, G) in Fig.~\ref{x-C1}. }
\begin{center}
\begin{tabular}{c c c c c c c c}
\hline\hline
Points &  $A$ & $B$ & $C$ &  $D$ & $E$ & $F$ & $G$\\
\hline 
$10^{10}C_1$ &  $0.0135$ & $0.535$ & $452$ &  $4.5\times10^{6}$ &$0.0175$   & $4.45$ & $94.5$\\
\hline
$10^{-5}\xi/\sqrt{\lambda}$ &  $0.38$ & $0.39$ & $0.69$ &  $2.58$ &$0.46$ & $0.52$ & $0.69$\\
\hline
$10^{7}Q$ &  $0.085$ & $1.542$ & $423.889$ &  $1.185\times10^6$ &$0.106$ & $9.390$ & $121.054$\\
\hline \hline
\end{tabular}
\end{center}\label{Points}
\end{table}

\begin{figure}[!thb]
  \centering
  \includegraphics[width=0.55\textwidth]{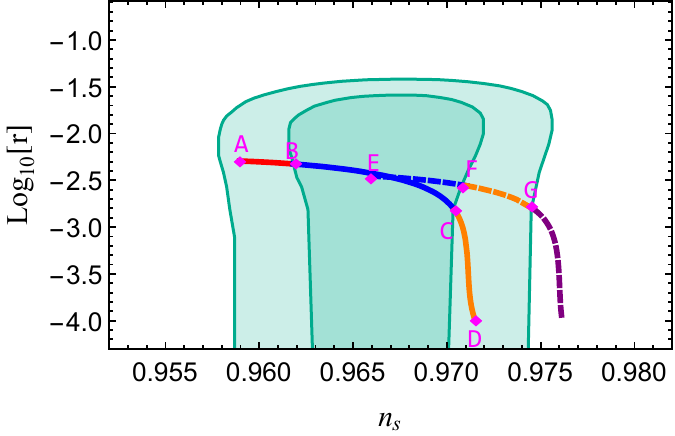}
  \caption{The projections of warm inflation in the $(n_s, r)$ planes are illustrated for both $N_e = 50$ and $N_e = 60$. The shaded regions represent the $1\sigma$ and $2\sigma$ experimental uncertainties from BAO, BICEP/KECK, and Planck data, with darker and lighter green denoting the respective confidence levels.~\cite{BICEP:2021xfz}.}\label{nsr}
\end{figure}

Upon substituting the model parameters depicted in Fig.~\ref{x-C1} into Eqs.~(\ref{r}) and (\ref{ns1}), we derive refined predictions for the model concerning $(n_s, r)$. Predictions associated with the Orange, Blue, and Purple model parameters in Fig.~\ref{x-C1} correspond to the respective colors in Fig.~\ref{nsr}, where solid lines represent $N_e = 50$ (dashed lines correspond to $N_e = 60$).

Notably, the predictions associated with the Orange and Blue fall within the $1\sigma$ and $2\sigma$ confidence intervals, respectively, while the Purple predictions extend beyond the defined experimental regions.
And the points A to G delineate the corresponding critical points. 

Examination of Fig.~\ref{nsr} reveals that, with the incorporation of Delta constraints, the predictions for $(n_s, r)$ under $N_e =50$ align with experimental constraints. Conversely, for $N_e = 60$, certain predictions for $(n_s, r)$ extend beyond the experimental limits, indicating a further refinement in constraining the model parameters.

\subsection{Evolution of the different quantities}

\begin{figure}[!thb]
  \centering
  \includegraphics[width=0.45\textwidth]{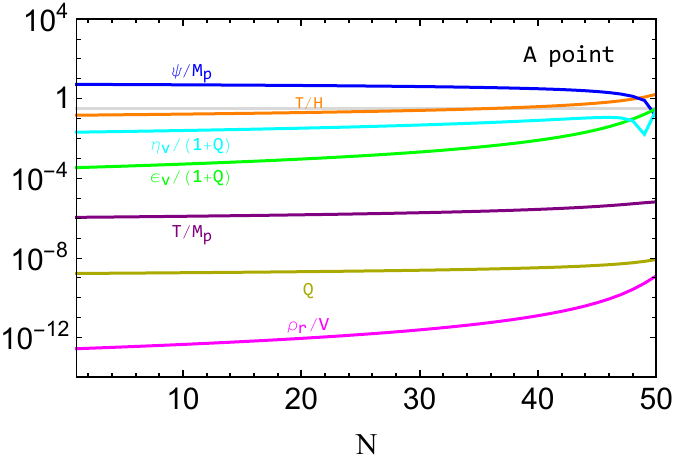}
  \includegraphics[width=0.45\textwidth]{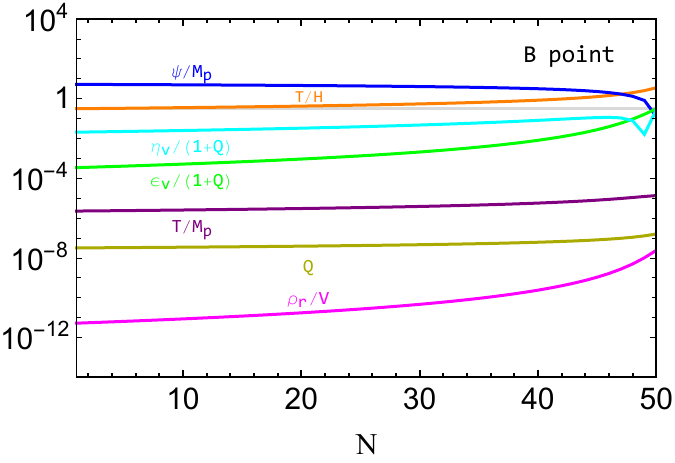}
  \includegraphics[width=0.45\textwidth]{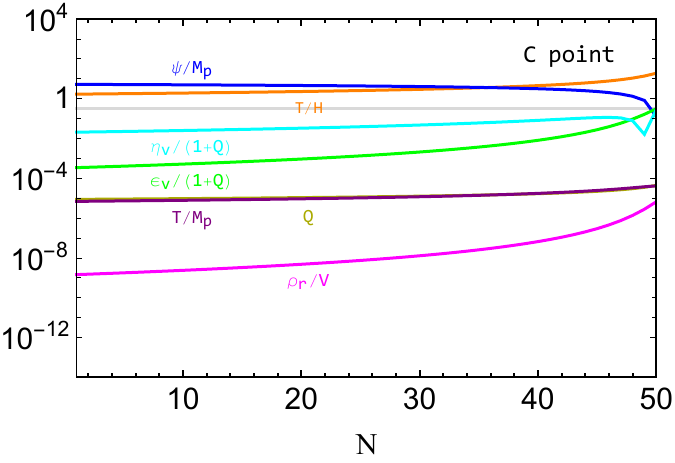}
  \includegraphics[width=0.45\textwidth]{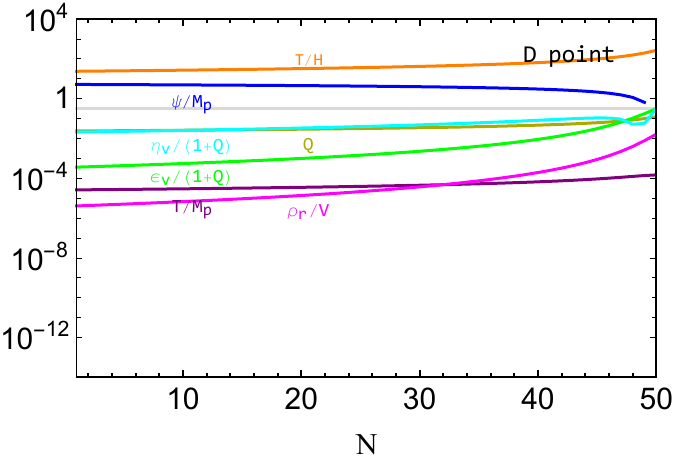}
  \includegraphics[width=0.45\textwidth]{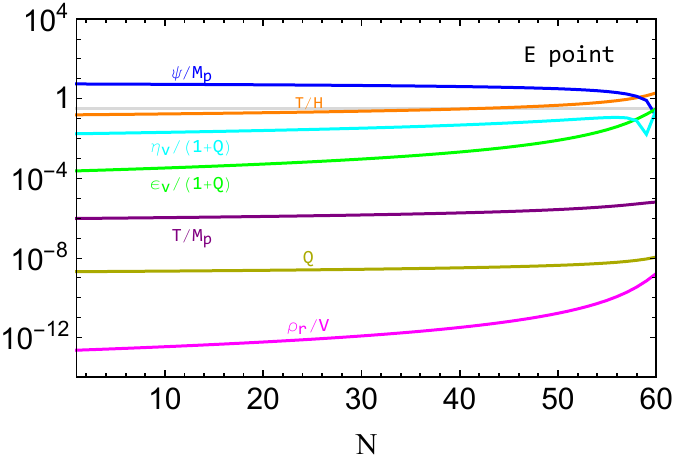}
  \includegraphics[width=0.45\textwidth]{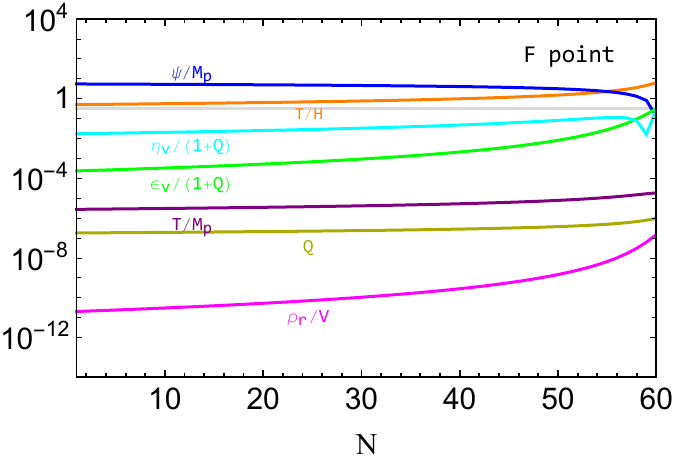}
  \includegraphics[width=0.45\textwidth]{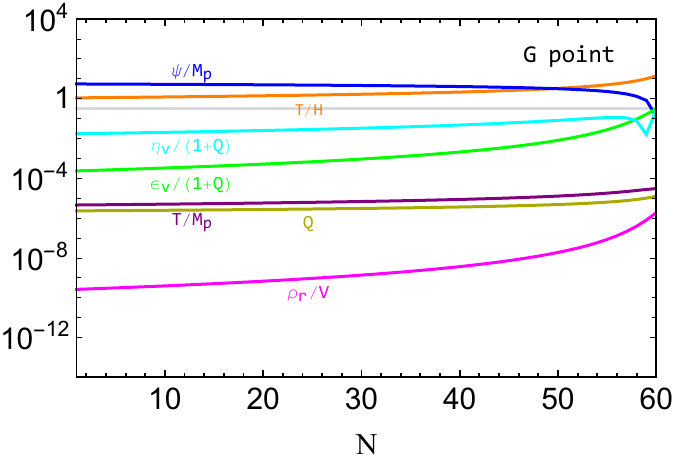}
  \caption{The evolution of various parameters in the WHI scenario for A, B, C, D, E, F and G points in the Tab.~\ref{x-C1}.}\label{XX-N}
\end{figure}

In Fig.~\ref{XX-N}, we employed the model parameter values from Tab.~\ref{Points} to calculate the evolution of various quantities with respect to $N$, representing the dynamics of these quantities during the inflationary process. It is noteworthy that the evolution trends of these quantities are similar in the plots corresponding to points A, B, C and D ($N_e=50$) and points E, F and G ($N_e=60$), with the only distinction being in their specific numerical values.

Throughout the entire inflationary process, an observation of the four plots for points A, B, C and D reveals that, as the model parameters $(C_1, \xi/\sqrt{\lambda})$ increase, the numerical value of $Q$ also rises correspondingly. However, overall, $Q$ remains at a relatively small level, i.e., $Q\ll1$. This pattern is similarly evident for points E, F and G. Consequently, WHI can be regarded as weak WHI.

As WHI falls under the category of weak WHI, discussions should refrain from employing the strong dissipation approximation. It is worth noting that despite the relatively small magnitude of $Q$, the relationship between $(\rho/ V, T/H, T/M_p)$ and $Q$ displayed in Fig.~\ref{XX-N} exhibits significant variations. Therefore, in the study of WHI, the impact of $Q$ should not be disregarded, even though it is relatively small.

\subsection{Primordial curvature power spectrum}

\begin{figure}[thb]
  \centering
  \includegraphics[width=0.45\textwidth]{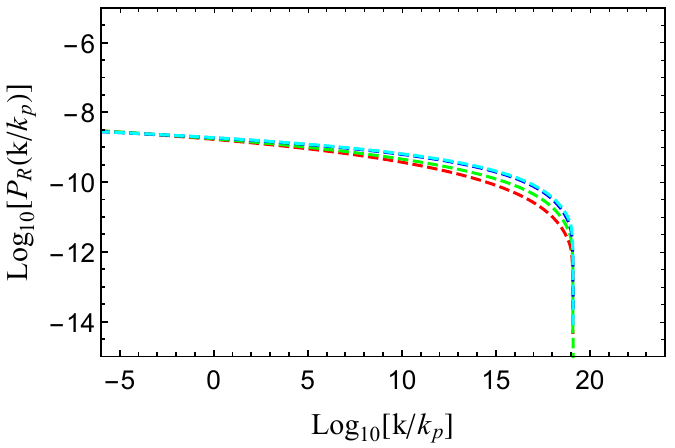}
  \includegraphics[width=0.45\textwidth]{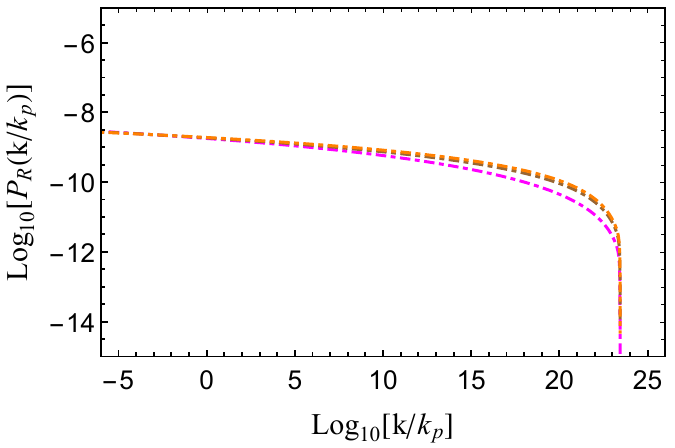}
  \caption{The primordial curvature power spectrum, $P_R(k/k_p)$ vs $\ln(k/k_p )$ is presented in the left panel for $N_e = 50$ and in the right panel for $N_e = 60$.}\label{kD}
\end{figure}

Utilizing the model parameter values from Tab.~\ref{Points} and employing Eq.~(\ref{spectrum}), we conducted an in-depth investigation of the primordial curvature power spectrum. We establish a connection between wave number $k$ and inflationary field $h$ by employing the relationship between $N_k$ and $k$ as given in Eq.~(\ref{Nk_k}), as well as the relationship between $N_k$ and the inflationary field $h$.
\begin{eqnarray}
	N_k=\ln\frac{a_e}{a_k}=\ln\frac{a_e}{a_p}+\ln\frac{a_p}{a_k}=N_p+\ln\bigg(\frac{k_p}{k }\frac{H_k}{H_p}\bigg)=N_p-\ln\frac{k}{k_p}+\ln\frac{H_k}{H_p}.\label{Nk_k}
\end{eqnarray}

 The variation of $\Delta_R(k/k_p)$ with $\ln(k/k_p )$ is illustrated in Fig.~\ref{kD}, where the dashed lines in red, green, blue, cyan, magenta, brown, and orange respectively correspond to the predictions associated with points A, B, C, D, E, F, and G.

It is noteworthy that the four dashed lines representing the $N_e=50$ scenarios almost completely overlap, as do the corresponding lines for $N_e=60$. This observation suggests that the model parameters $(C_1, \xi/\sqrt{\lambda})$ have a relatively minor impact on $P_R(k)$. Given that $(C_1, \xi/\sqrt{\lambda})$ can be translated into $Q$, this also implies that $Q$ exerts a relatively small influence on $P_R(k)$.

\subsection{Relic gravitational wave spectrum}

\begin{figure}[thb]
  \centering
  \includegraphics[width=0.45\textwidth]{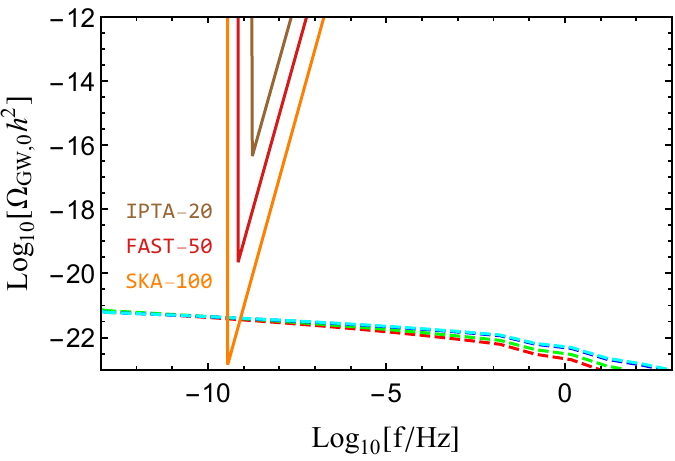}
  \includegraphics[width=0.45\textwidth]{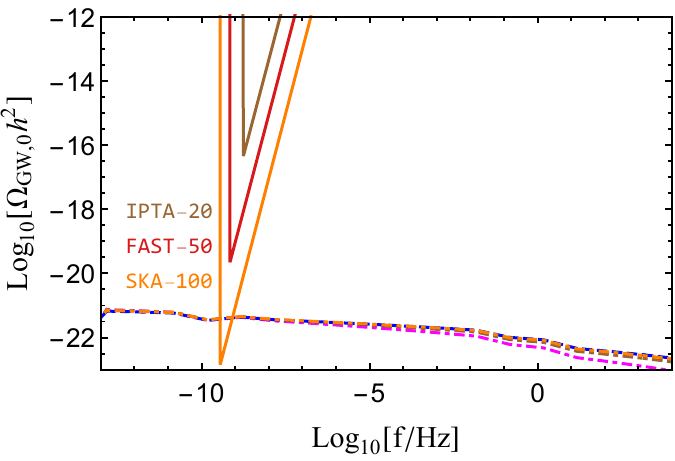}
  \caption{The spectrum of secondary gravitational waves at the present time, generated from WHI as a function of the frequency of GWs~\cite{Carilli:2004nx,Nan:2011um,IPTA}.}\label{GW}
\end{figure}

Fig.~\ref{GW} displays the spectrum of secondary gravitational waves at the present time. The dashed lines represent our predicted results, and for the four scenarios with $N_e=50$, the differences in predicted gravitational waves are relatively small. Similarly, the three scenarios with $N_e=60$ exhibit comparable trends, primarily because the corresponding variations in Delta are modest, ultimately resulting in a limited influence of $Q$ on gravitational waves. Despite the relatively modest amplitudes predicted by the model, there is potential for validation through future experiments, such as the SKA100, especially around the frequency $f\approx 10^{-9.5}~\rm{Hz}$.

\section{Summary}\label{sec:Sum}

In this study, we systematically investigated the dependencies of various physical quantities on $Q$ during the WHI process. Recognizing the pivotal role of $Q$ in WHI, we initially derived a comprehensive relationship between each physical quantity in WHI and $Q$, avoiding any a priori assumptions regarding strong or weak dissipation.

Subsequently, accounting for the amplitude of scalar fluctuations' constraint, we determined the feasible parameter space ($C_1,\xi/\sqrt{\lambda}$). Further predictions of $r$ and $n_s$ were made and compared with the CMB observational results. We found that, for the scenario $N_e=50$, the predictions completely conformed to the constraints imposed by the CMB, while for the scenario $N_e=60$, the predictions partially aligned within the CMB constraints.

Upon further analyzing the variation of the dissipation coefficient with $N$, we observed that although $Q$ exhibited significant changes with variations in model parameters ($C_1,\xi/\sqrt{\lambda}$), overall, Q remained at extremely low levels, i.e., $Q\ll1$. This indicates that Higgs warm inflation can be categorized as weakly dissipative warm inflation. Despite the small numerical value of $Q$, its dependence on various physical quantities under the influence of parameters ($C_1,\xi/\sqrt{\lambda}$), especially in the $N_e=60$ scenario, is pronounced and should not be overlooked. This is clearly illustrated in Fig.~\ref{XX-N}.

Finally, we conducted an in-depth investigation into curvature perturbations during WHI process and found that $Q$ had a negligible impact on them, further confirming the model's prediction of GWs variations. This conclusion is also well-reflected in Fig.~\ref{GW}, which additionally suggest that GWs generated by WHI are expected to be validated in future observational experiments, particularly through the detection capabilities of the SKA100 experiment. This discovery not only strengthens our theoretical understanding of the model but also provides crucial directions and expectations for future experiments.

\acknowledgments

Wei Cheng was supported by Chongqing Natural Science Foundation project under Grant No. CSTB2022NSCQ-MSX0432, by Science and Technology Research Project of Chongqing Education Commission under Grant No. KJQN202200621 and No. KJQN202200605, and by Chongqing Human Resources and Social Security Administration Program under Grants No. D63012022005. Xue-Wen Chen was supported by Natural Science Foundation of Chongqing No. cstc2021jcyjmsxmX0678, and  by Science and Technology Research Program of Chongqing Municipal Education Commission No. KJQN202201527. This work was also supported in part by the Fundamental Research Funds for the Central Universities under Grant No. 2021CDJQY-011,
and by the National Natural Science Foundation of China under Grant  No. 12147102.
Ruiyu Zhou was supported by supported by the National Natural Science Foundation of China under Grant No. 12305109, Chongqing Natural Science Foundation project under Grant No. CSTB2022NSCQ-MSX0534, and by Science and Technology Research Project of Chongqing Municipal Education Commission under Grant No.KJQN202300614.

\section{Appendix}\label{App}

For the WHI, the more detailed information for each component in the scalar power spectrum is provided below:

For the investigation of the front factor ($\frac{H^{2}}{2\pi {\dot h}}$), the potential energy (Eq.~\ref{potential}) serves to derive $H$, given by:
\begin{eqnarray}
H^{2}=\frac{\lambda  M_p^2}{12 \xi ^2 \left(e^{-\frac{\sqrt{\frac{2}{3}} h }{M_p}}+1\right)^2}\,.\label{H2}
\end{eqnarray}
Simultaneously, $\dot h$ can be calculated using Eq.~\ref{eominfh}, expressed as:
\begin{eqnarray}
{\dot h}\approx -\frac{U'(h)}{3 (Q+1) H}=-\frac{\sqrt{2} M_p e^{\frac{\sqrt{\frac{2}{3}} h }{M_p}} \sqrt{\frac{\lambda  M_p^2}{\xi ^2}}}{3 (Q+1) \left(e^{\frac{\sqrt{\frac{2}{3}} h }{M_p}}+1\right)^2}\,.\label{H2psi}
\end{eqnarray}
Combining Eqs.~(\ref{H2}) and (\ref{H2psi}), the front factor can ultimately be represented as follows:
\begin{eqnarray}
\frac{H^{2}}{2\pi {\dot h}}=-\frac{(Q+1) e^{\frac{\sqrt{\frac{2}{3}} h }{M_p}} \sqrt{\frac{\lambda  M_p^2}{\xi ^2}}}{8 \sqrt{2} \pi  M_p}\,.\label{H2psid}
\end{eqnarray}

Concerning the modified factor (${\cal{F}}$) in Eq.~\ref{spectrum},
\begin{eqnarray}
{\cal{F}} = \left( 1 + 2n_k +\left(\frac{T_k}{H_k}\right)\frac{2\sqrt{3}\,\pi\,Q_k}{\sqrt{3+4\pi\,Q_k}}\right)G(Q_k)\,,
\label{F}
\end{eqnarray}
where $n$ represents the Bose-Einstein distribution function with a specific expression $n_k = 1/\big( \exp{H_k/T_k} - 1 \big)$, and $T$ denotes the temperature of the thermal bath. The temperature is intricately linked to the dissipative coefficient $Q$ and the Hubble constant $H$. By solving the equation for the dissipative coefficient, $Q=\Gamma/3H=C_{1}T/3H$, we can derive:
\begin{eqnarray}
T=\frac{1}{6^{1/4}}\left(\frac{\lambda  Q M_p^4 e^{\frac{2 \sqrt{\frac{2}{3}} h }{M_p}}}{C_{r} \xi ^2 (Q+1)^2 \left(e^{\frac{\sqrt{\frac{2}{3}} h }{M_p}}+1\right)^4}\right)^{1/4}\,.\label{Tex}
\end{eqnarray}

Dividing the above relation with $H$, we obtain:
\begin{eqnarray}
\frac{T}{H}=2^{3/4}\,3^{1/4}\left(\frac{\lambda  M_p^2}{\xi ^2 \left(e^{-\frac{\sqrt{\frac{2}{3}} h }{M_p}}+1\right)^2}\right)^{-1/2}\left(\frac{\lambda  Q M_p^4 e^{\frac{2 \sqrt{\frac{2}{3}} h }{M_p}}}{\text{Cr} \xi ^2 (Q+1)^2 \left(e^{\frac{\sqrt{\frac{2}{3}} h }{M_p}}+1\right)^4}\right)^{1/4}\,.\label{TH}
\end{eqnarray}

In a thermal bath, the interaction between the inflaton and radiation is characterized by the growth factor $G(Q_k)$. This function was introduced by Graham et al.\cite{Graham:2009bf}, with subsequent discussions in studies\cite{Bastero-Gil:2011rva,Bastero-Gil:2018uep}. The evolution of $G(Q_k)$ depends on $\Gamma$. In this work, we consider $\Gamma \propto T$, yielding the expression for $G(Q_k)$ in \cite{Benetti:2016jhf,Bastero-Gil:2018uep}:
\begin{eqnarray}
G(Q_k)_{\rm linear} = 1+0.0185Q^{2.315}_{k}+0.335Q^{1.364}_{k}
\,. \label{ga}
\end{eqnarray}

In this study, we have observed the phenomenon of $Q\ll 1$ during the WHIary epoch. Clearly, for small values of $Q$, i.e., $Q\ll 1$, it is noteworthy that the modified scalar power spectrum exhibits no significant variation with changes in $Q$.

Through a comprehensive analysis of the aforementioned factors, it is evident that all these factors can be expressed in terms of model parameters the non-minimal coupling $\xi$, the coupling $\lambda$, the field $h$, and the dissipation coefficient $Q$. Subsequently, based on the definition of the dissipation coefficient $Q=\Gamma/3H=C_{1}T/3H$ and in conjunction with the Eq.~(\ref{Tex}), the dissipation coefficient $Q$ is formulated in terms of $\xi$, $\lambda$, and $h$:

\begin{eqnarray}
e^{\frac{\sqrt{\frac{2}{3}} h }{M_p}}\approx \frac{2 \sqrt{\frac{2}{3}} C_1^2 \xi }{3 \sqrt{\frac{g_\star \pi^2}{30}} \sqrt{\lambda } Q^{5/2}}\quad\rightarrow\quad Q=\frac{\left(\frac{2}{3}\right)^{3/5} C_1^{4/5} \xi^{2/5} }{\left(\frac{g_\star \pi^2}{30}\right)^{1/5} \lambda^{1/5}}e^{-\frac{2 \sqrt{\frac{2}{3}} h}{5 M_p}}\,.\label{phiex}
\end{eqnarray}

Therefore, the scalar power spectrum is entirely characterized by the model parameters $\xi$, $\lambda$, and the field $h$.

\bibliography{lit}


%
%

\end{document}